\documentclass[12pt,onecolumn,draftclsnofoot]{IEEEtran}
\usepackage{fancyhdr}
\usepackage{amsmath,epsfig}
\usepackage{threeparttable}
\usepackage{epsf,epsfig}
\usepackage{amsmath}
\usepackage{amssymb}
\usepackage{amsfonts}
\usepackage{algorithmic}
\usepackage{algorithm}
\usepackage[noadjust]{cite}
\usepackage{dsfont}
\usepackage{subfigure}

\begin{document}
\newtheorem{theorem}{\bf Theorem}
\newtheorem{acknowledgement}[theorem]{Acknowledgement}
\newtheorem{axiom}[theorem]{Axiom}
\newtheorem{case}[theorem]{Case}
\newtheorem{claim}[theorem]{Claim}
\newtheorem{conclusion}[theorem]{Conclusion}
\newtheorem{condition}[theorem]{Condition}
\newtheorem{conjecture}[theorem]{Conjecture}
\newtheorem{criterion}[theorem]{Criterion}
\newtheorem{definition}[theorem]{Definition}
\newtheorem{example}[theorem]{Example}
\newtheorem{exercise}[theorem]{Exercise}
\newtheorem{lemma}{\bf Lemma}
\newtheorem{corollary}{\bf Corrollary}
\newtheorem{notation}[theorem]{Notation}
\newtheorem{problem}[theorem]{Problem}
\newtheorem{proposition}{\bf Proposition}
\newtheorem{solution}[theorem]{Solution}
\newtheorem{summary}[theorem]{Summary}
\newtheorem{assumption}{\bf AS}
\newtheorem{examp}{\bf Example}
\newtheorem{probform}{\bf Problem}
\def\remark{{\noindent \bf Remark:\hspace{0.5em}}}

\def\qed{$\Box$}
\def\QED{\mbox{\phantom{m}}\nolinebreak\hfill$\,\Box$}
\def\proof{\noindent{\bf Proof: }}
\def\poof{\noindent{\bf Sketch of Proof: }}
\def
\endproof{\hspace*{\fill}~\qed
\par
\endtrivlist\unskip}
\def\endproof{\hspace*{\fill}~\qed\par\endtrivlist\vskip3pt}

\def\eps{\varepsilon}
\def\phi{\varphi}
\def\Lsp{{\boldsymbol L}}
\def\Bsp{{\boldsymbol B}}
\def\lsp{{\boldsymbol\ell}}
\def\Ltsp{{\Lsp^2}}
\def\Lpsp{{\Lsp^p}}
\def\Linsp{{\Lsp^{\infty}}}
\def\LtR{{\Lsp^2(\Rst)}}
\def\ltZ{{\lsp^2(\Zst)}}
\def\ltsp{{\lsp^2}}
\def\ltZt{{\lsp^2(\Zst^{2})}}
\def\ninN{{n{\in}\Nst}}
\def\oh{{\frac{1}{2}}}
\def\grass{{\cal G}}
\def\ord{{\cal O}}
\def\dist{{d_G}}
\def\conj#1{{\overline#1}}
\def\ntoinf{{n \rightarrow \infty }}
\def\toinf{{\rightarrow \infty }}
\def\tozero{{\rightarrow 0 }}
\def\trace{{\operatorname{trace}}}
\def\ord{{\cal O}}
\def\UU{{\cal U}}
\def\rank{{\operatorname{rank}}}
\def\acos{{\operatorname{acos}}}

\def\SINR{\textrm{SINR}}
\def\SNR{\textrm{SNR}}
\def\SIR{\textrm{SIR}}

\setcounter{page}{1}

\newcommand{\eref}[1]{(\ref{#1})}
\newcommand{\fig}[1]{Fig.\ \ref{#1}}

\def\bydef{:=}
\def\ba{{\mathbf{a}}}
\def\bb{{\mathbf{b}}}
\def\bc{{\mathbf{c}}}
\def\bd{{\mathbf{d}}}
\def\bee{{\mathbf{e}}}
\def\bff{{\mathbf{f}}}
\def\bg{{\mathbf{g}}}
\def\bh{{\mathbf{h}}}
\def\bi{{\mathbf{i}}}
\def\bj{{\mathbf{j}}}
\def\bk{{\mathbf{k}}}
\def\bl{{\mathbf{l}}}
\def\bm{{\mathbf{m}}}
\def\bn{{\mathbf{n}}}
\def\bo{{\mathbf{o}}}
\def\bp{{\mathbf{p}}}
\def\bq{{\mathbf{q}}}
\def\br{{\mathbf{r}}}
\def\bs{{\mathbf{s}}}
\def\bt{{\mathbf{t}}}
\def\bu{{\mathbf{u}}}
\def\bv{{\mathbf{v}}}
\def\bw{{\mathbf{w}}}
\def\bx{{\mathbf{x}}}
\def\by{{\mathbf{y}}}
\def\bz{{\mathbf{z}}}
\def\b0{{\mathbf{0}}}

\def\bA{{\mathbf{A}}}
\def\bB{{\mathbf{B}}}
\def\bC{{\mathbf{C}}}
\def\bD{{\mathbf{D}}}
\def\bE{{\mathbf{E}}}
\def\bF{{\mathbf{F}}}
\def\bG{{\mathbf{G}}}
\def\bH{{\mathbf{H}}}
\def\bI{{\mathbf{I}}}
\def\bJ{{\mathbf{J}}}
\def\bK{{\mathbf{K}}}
\def\bL{{\mathbf{L}}}
\def\bM{{\mathbf{M}}}
\def\bN{{\mathbf{N}}}
\def\bO{{\mathbf{O}}}
\def\bP{{\mathbf{P}}}
\def\bQ{{\mathbf{Q}}}
\def\bR{{\mathbf{R}}}
\def\bS{{\mathbf{S}}}
\def\bT{{\mathbf{T}}}
\def\bU{{\mathbf{U}}}
\def\bV{{\mathbf{V}}}
\def\bW{{\mathbf{W}}}
\def\bX{{\mathbf{X}}}
\def\bY{{\mathbf{Y}}}
\def\bZ{{\mathbf{Z}}}

\def\mA{{\mathbb{A}}}
\def\mB{{\mathbb{B}}}
\def\mC{{\mathbb{C}}}
\def\mD{{\mathbb{D}}}
\def\mE{{\mathbb{E}}}
\def\mF{{\mathbb{F}}}
\def\mG{{\mathbb{G}}}
\def\mH{{\mathbb{H}}}
\def\mI{{\mathbb{I}}}
\def\mJ{{\mathbb{J}}}
\def\mK{{\mathbb{K}}}
\def\mL{{\mathbb{L}}}
\def\mM{{\mathbb{M}}}
\def\mN{{\mathbb{N}}}
\def\mO{{\mathbb{O}}}
\def\mP{{\mathbb{P}}}
\def\mQ{{\mathbb{Q}}}
\def\mR{{\mathbb{R}}}
\def\mS{{\mathbb{S}}}
\def\mT{{\mathbb{T}}}
\def\mU{{\mathbb{U}}}
\def\mV{{\mathbb{V}}}
\def\mW{{\mathbb{W}}}
\def\mX{{\mathbb{X}}}
\def\mY{{\mathbb{Y}}}
\def\mZ{{\mathbb{Z}}}

\def\cA{\mathcal{A}}
\def\cB{\mathcal{B}}
\def\cC{\mathcal{C}}
\def\cD{\mathcal{D}}
\def\cE{\mathcal{E}}
\def\cF{\mathcal{F}}
\def\cG{\mathcal{G}}
\def\cH{\mathcal{H}}
\def\cI{\mathcal{I}}
\def\cJ{\mathcal{J}}
\def\cK{\mathcal{K}}
\def\cL{\mathcal{L}}
\def\cM{\mathcal{M}}
\def\cN{\mathcal{N}}
\def\cO{\mathcal{O}}
\def\cP{\mathcal{P}}
\def\cQ{\mathcal{Q}}
\def\cR{\mathcal{R}}
\def\cS{\mathcal{S}}
\def\cT{\mathcal{T}}
\def\cU{\mathcal{U}}
\def\cV{\mathcal{V}}
\def\cW{\mathcal{W}}
\def\cX{\mathcal{X}}
\def\cY{\mathcal{Y}}
\def\cZ{\mathcal{Z}}
\def\cd{\mathcal{d}}
\def\Mt{M_{t}}
\def\Mr{M_{r}}
\def\O{\Omega_{M_{t}}}
\renewcommand{\eqref}[1]{(\ref{#1})}
\newcommand{\figref}[1]{{Fig.}~\ref{#1}}
\newcommand{\tabref}[1]{{Table}~\ref{#1}}


\renewcommand{\mod}{\tx{mod}}
\newcommand{\m}[1]{\mathbf{#1}}
\newcommand{\td}[1]{\tilde{#1}}
\newcommand{\sbf}[1]{\scriptsize{\textbf{#1}}}
\newcommand{\stxt}[1]{\scriptsize{\textrm{#1}}}
\newcommand{\suml}[2]{\sum\limits_{#1}^{#2}}
\newcommand{\sumlk}{\sum\limits_{k=0}^{K-1}}
\newcommand{\eqhsp}{\hspace{10 pt}}
\newcommand{\tx}[1]{\texttt{#1}}
\newcommand{\Hz}{\ \tx{Hz}}
\newcommand{\sinc}{\tx{sinc}}
\newcommand{\tr}{\tx{tr}}
\newcommand{\diag}{\tx{diag}}
\newcommand{\MAI}{\tx{MAI}}
\newcommand{\ISI}{\tx{ISI}}
\newcommand{\IBI}{\tx{IBI}}
\newcommand{\CN}{\tx{CN}}
\newcommand{\CP}{\tx{CP}}
\newcommand{\ZP}{\tx{ZP}}
\newcommand{\ZF}{\tx{ZF}}
\newcommand{\SP}{\tx{SP}}
\newcommand{\MMSE}{\tx{MMSE}}
\newcommand{\MINF}{\tx{MINF}}
\newcommand{\RC}{\tx{MP}}
\newcommand{\MBER}{\tx{MBER}}
\newcommand{\MSNR}{\tx{MSNR}}
\newcommand{\MCAP}{\tx{MCAP}}
\newcommand{\vol}{\tx{vol}}
\newcommand{\ah}{\hat{g}}
\newcommand{\tg}{\tilde{g}}
\newcommand{\teta}{\tilde{\eta}}
\newcommand{\heta}{\hat{\eta}}
\newcommand{\uh}{\m{\hat{s}}}
\newcommand{\eh}{\m{\hat{\eta}}}
\newcommand{\hv}{\m{h}}
\newcommand{\hh}{\m{\hat{h}}}
\newcommand{\Po}{P_{\textrm{out}}}
\newcommand{\Poh}{\hat{P}_{\textrm{out}}}
\newcommand{\Ph}{\hat{\gamma}}
\newcommand{\mat}[1]{\begin{matrix}#1\end{matrix}}
\newcommand{\ud}{^{\dagger}}
\newcommand{\C}{\mathcal{C}}
\newcommand{\nn}{\nonumber}
\newcommand{\nInf}{U\rightarrow \infty}

\title{\setlength{\baselineskip}{40pt} Space Division Multiple Access with a Sum Feedback Rate Constraint}
\author{\large \setlength{\baselineskip}{20pt}Kaibin Huang, Robert W. Heath,
Jr and Jeffrey G. Andrews
\thanks{\setlength{\baselineskip}{10pt} Authors are with The University of Texas at Austin. Email: \{khuang,
rheath, jandrews \}@ece.utexas.edu}
\thanks{\setlength{\baselineskip}{10pt} This work is funded by
Freescale Inc. and the National Science Foundation under grants
CCF-514194. }}\maketitle
 \vspace{-60pt}
\begin{abstract}\setlength{\baselineskip}{15pt}
On a multi-antenna broadcast channel, simultaneous transmission to
multiple users by joint beamforming and scheduling is capable of
achieving high throughput, which grows double logarithmically with
the number of users. The sum rate for channel state information
(CSI) feedback, however, increases linearly with the number of
users, reducing the effective uplink capacity. To address this
problem, a novel space division multiple access (SDMA) design is
proposed, where the sum feedback rate is upper-bounded by a
constant. This design consists of algorithms for CSI quantization,
threshold based CSI feedback, and joint beamforming and scheduling.
The key feature of the proposed approach is the use of feedback
thresholds to select feedback users with large channel gains and
small CSI quantization errors such that the sum feedback rate
constraint is satisfied. Despite this constraint, the proposed SDMA
design is shown to achieve a sum capacity growth rate close to the
optimal one. Moreover, the feedback overflow probability for this
design is found to decrease exponentially with the difference
between the allowable and the average sum feedback rates. Numerical
results show that the proposed SDMA design is capable of attaining
higher sum capacities than existing ones, even though the sum
feedback rate is bounded.

\end{abstract}

\section{Introduction}\label{Section:Intro}
For a multi-antenna communication downlink, \emph{space division
multiple access} (SDMA) allows simultaneous transmission through the
spatial separation of scheduled users. The high throughput of SDMA
led to its inclusion in the IEEE 802.16e standard
\cite{IEEE802-16e}. Compared with the optimal SDMA strategy that
uses \emph{dirty paper coding}
\cite{Costa:WriteDirtyPaper:83,Caire:AchivThroghputBroadcastChan:2003},
SDMA with transmit beamforming has suboptimal performance but a
low-complexity transmitter. Various methods for designing SDMA under
beamforming constraints have been proposed recently, including zero
forcing
\cite{Choi:TXPreprocMuMIMO:2004,Wong:ChanDiagMuMIMO:2003,SpencerSwindleETAL:ZFsdma:2004,Dimic:LowCompDLBeamfMaxCap:2004},
a signal-to-interference-plus-noise-ratio (SINR) constraint
\cite{SchubertBoche:solsdma:04},
 minimum mean squared error (MMSE) \cite{Serbetli:TxRxOptimMuMIMO:2004},
and channel decomposition \cite{Choi:ComplementBeamf:2006}. In a
network, SDMA beamforming can be combined with scheduling to further
improve the throughput by exploiting multi-user diversity, which
refers to the selection of users with good channels for transmission
\cite{Knopp:InfoCapPwrCtrlMu:1995,SwannackWorenell:BroadcastChanLimitedCSI:2005,
YooGoldsmith:OptimBroadcastZeroForcingBeam:2006,
YooJindal:FiniteRateBroadcastMUDiv:2006,ChoiForenza:OppSDMABeamSel:06,SharifHassibi:CapMIMOBroadcastPartSideInfo:Feb:05,
ShenChen:LowCompUsrSelBD:2006}. Typically, joint beamforming and
scheduling for SDMA requires users to send back their channel state
information (CSI). Therefore, given that all users share a common
feedback channel, the sum feedback rate can rapidly become a
bottleneck for a SDMA system with a large number users. That
motivates us to address in  this paper the following questions:
\emph{How to design a SDMA downlink with a bounded sum feedback
rate? Does this sum feedback rate constraint significantly affect
the system performance?}

\subsection{Prior Work and Motivation}
The sum feedback rate of a downlink system can be reduced by
applying a feedback threshold, where users below the threshold do
not send back CSI. This feedback reduction algorithm was first
proposed in \cite{Gesbert:HowMuchFbMUDivWorth:2004} for a downlink
system with single-input-single-output (SISO) channels, where only
users meeting a signal-to-noise-ratio (SNR) threshold are allowed to
send back SNR information for scheduling. This algorithm is shown to
reduce the sum feedback rate significantly. To further reduce the
sum feedback rate, the feedback reduction algorithm in
\cite{Gesbert:HowMuchFbMUDivWorth:2004} is modified in
\cite{Hassel:MUDivMulFBThresholds:2005} to have an adaptive
threshold. The drawback of this modified algorithm is the feedback
delay due to multiple rounds of feedback and also the additional
feedback cost incurred by this process. In
\cite{Sanayei:MUDiv1BitFb:2005,Sanayei:OppBeamLimitedFb:2005}, for
both SISO and multiple-input-single-output (MISO) channels,
combining a feedback threshold and one bit feedback per user is
shown to achieve the optimal growth rate of the sum capacity with
the number of users. A common problem shared by these feedback
reduction algorithms is that the sum feedback rate increases
linearly with the number of users, placing a burden on the uplink
channel if the number of users is large.

To constrain the sum feedback rate,  an approach combining a
feedback threshold and contention feedback is proposed in
\cite{TangHeath:OppFbDLMuDiv:2005} for SISO channels, where feedback
users contend for the use of a common feedback channel. By extending
this approach to MIMO channels, a SDMA algorithm is proposed in
\cite{TangHeath:OppFbMuMIMOLinearRx:2005}, which nevertheless has
limitations for practical implementation. First, the number of
simultaneous users supported by space division is limited by the
number of receive antennas for each user, which is usually very
small. Second, every user must inefficiently perform zero-forcing
equalization even though only a small subset of users is scheduled
for transmission. These limitations motivate us to consider a more
practical downlink system.

In the literature of SDMA with transmit beamforming, the sum
feedback rate constraint has not been considered as most work
focuses on feedback reduction for individual users. For the
\emph{opportunistic SDMA} (OSDMA) algorithm proposed in
\cite{SharifHassibi:CapMIMOBroadcastPartSideInfo:Feb:05}, the
feedback of each user is reduced to a few bits by constraining the
choice of a beamforming vector to a set of orthogonal vectors. The
sum capacity of OSDMA can be increased by selecting orthogonal
beamforming vectors from multiple sets of orthogonal vectors, which
motivates the \emph{OSDMA with beam selection} (OSDMA-BS)
\cite{ChoiForenza:OppSDMABeamSel:06} and the \emph{OSDMA with
limited feedback} (OSDMA-LF) \cite{Huang:JointBeamScheduleLimtFb:06}
algorithms\footnote{Limited feedback refers to quantization and
feedback of CSI
\cite{LovHeaETAL:WhatValuLimiFee:Oct:2004,LoveHeathBook}}. These two
algorithms assign beamforming vectors at mobiles and the base
station, respectively. Existing SDMA algorithms share the drawback
of having a sum feedback rate that increases linearly with the
number of users. This motivates us to apply a sum feedback rate
constraint on SDMA.

\subsection{Contributions and Organization}
We propose an algorithm for a SDMA downlink with orthogonal
beamforming and the average sum rate for CSI feedback upper-bounded
by a constant, which is referred to as the sum feedback rate
constraint. This constraint is enforced by using two feedback
thresholds for selecting feedback users, which gives the name of the
algorithm: \emph{OSDMA with threshold feedback} (OSDMA-TF). First, a
feedback threshold on users' channel power selects users with large
channel gains for feedback. Second, a threshold on users' channel
quantization errors prevents CSI quantization from stopping the
growth of the sum capacity with the number of users
\cite{Huang:JointBeamScheduleLimtFb:06,YooJindal:FiniteRateBroadcastMUDiv:2006}.
The key differences between OSDMA-TF and existing algorithms are
summarized as follows. Contrary to the sum feedback reduction
algorithms for SISO channels
\cite{Gesbert:HowMuchFbMUDivWorth:2004,YangAlouini:FurtherResultSelMuDiv:2004,Hassel:MUDivMulFBThresholds:2005,
Sanayei:MUDiv1BitFb:2005,Sanayei:OppBeamLimitedFb:2005} and other
OSDMA algorithms with finite-rate feedback
\cite{SharifHassibi:CapMIMOBroadcastPartSideInfo:Feb:05,ChoiForenza:OppSDMABeamSel:06,Huang:JointBeamScheduleLimtFb:06},
OSDMA-TF satisfies the sum feedback rate constraint. Among downlink
algorithms enforcing this constraint, OSDMA-TF has the advantage of
supporting simultaneous users compared with the SISO contention
feedback algorithm in \cite{TangHeath:OppFbDLMuDiv:2005} and the
advantage of having simple receivers for subscribers compared with
the MIMO contention feedback algorithm in
\cite{TangHeath:OppFbMuMIMOLinearRx:2005}.

The main contributions of this paper are the OSDMA-TF algorithm, the
design of feedback thresholds for enforcing the sum feedback rate
constraint, and the analysis of the impact of this constraint on the
sum capacity. First, the OSDMA-TF sub-algorithms for \emph{CSI
quantization at users}, \emph{selection of feedback users using
thresholds} and \emph{joint beamforming and scheduling at a base
station} are proposed. Second, the feedback thresholds on users'
channel power and channel quantization errors are designed such that
the sum feedback rate constraint is satisifed. Third, from an
upper-bound, the feedback overflow probability is found to decrease
approximately exponentially with the difference between the
allowable and the average sum feedback rates. Fourth, it is shown
that  the  growth rate of the sum capacity with the number of users
can be made arbitrarily close to the optimal one by having a
sufficiently large sum feedback rate. Last, OSDMA-TF is compared
with several existing SDMA algorithms and is found to be capable of
achieving higher sum capacities despite the sum feedback rate
constraint. The main conclusion of this paper is that the proposed
SDMA algorithm allows a sum feedback rate constraint to be applied
on a SDMA downlink without causing any appreciable negative impact.

The remainder of this paper is organized as follows. The system
model is described in Section~\ref{Section:Sys}. The algorithms used
for CSI quantization, CSI feedback and joint beamforming and
scheduling are presented in Section~\ref{Section:Algo}. The feedback
thresholds are derived in Section~\ref{Section:FbTh}, along with an
upper bound for the feedback overflow probability. In
Section~\ref{Section:CapAna}, the sum capacity for OSDMA-TF is
analyzed.  The performance of OSDMA-TF is compared with existing
SDMA algorithms using Monte Carlo simulations in
Section~\ref{Section:Numerical}, followed by concluding remarks in
Section~\ref{Section:Conclusion}.

\section{System Model} \label{Section:Sys}
\begin{figure}[h]
\centering
\includegraphics[width=13cm]{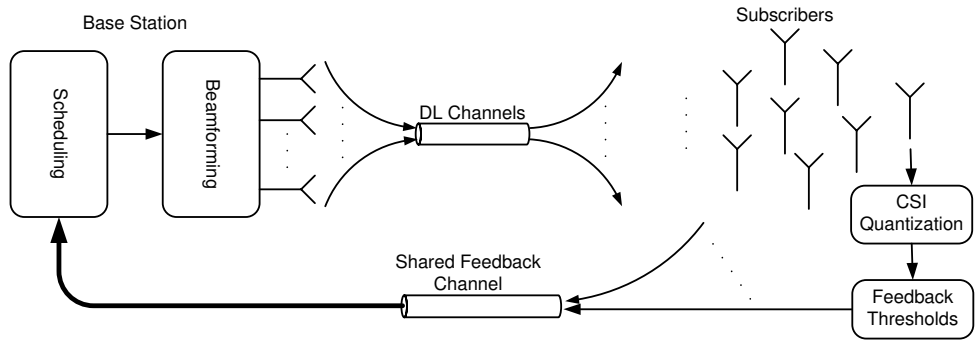}\\
\caption{SDMA Downlink system with feedback thresholds
}\label{Fig:DLSYS}
\end{figure}

The downlink system illustrated in Fig.~\ref{Fig:DLSYS} is described
as follows. A base station with $N_t$ antennas transmits data
simultaneously to $N_t$ scheduled users chosen from a total of $U$
users, each with  one receive antenna. The base station separates
the multi-user data streams by using beamforming, i.e. assigning a
beamforming vector to each of the $N_t$ scheduled users. The
beamforming vectors $\{\bw_n\}_{n=1}^{N_t}$ are selected from
multiple sets of unitary orthogonal vectors following the beam and
scheduling algorithm described in Section~\ref{Section:BeamSel}. The
received signal of the $n$th scheduled user is expressed as
\begin{equation}\label{Eq:Sys}
    y_n = \sqrt{P}\suml{i=1}{N_t}\bh^\dagger_n\bw_ix_i + \nu_n, \quad n=1,\cdots, N_t,
\end{equation}
where we use the following notation
  \begin{description}
  \item[$N_t$] number of transmit antennas and also number of
  scheduled users;
  \item[$\bh_n$]  ($N_t\times 1$ vector) downlink channel;
  \item[$P$] transmit power;
  \item[$\bw_n$] ($N_t\times 1$ vector) beamforming vector with
  $\|\bw_n\|^2=1$;
  \item[$x_n$] transmitted symbol with $|x_n| = 1$;
  \item[$y_n$] received symbol; and
  \item[$\nu_n$] AWGN sample with $\nu_n\in\mathcal{CN}(0,1)$.
\end{description}

We assume that each user quantizes his/her CSI and sends it back
following a feedback algorithm to be discussed in
Section~\ref{Section:FBScheme}. Furthermore, all users share a
common feedback channel. Therefore, it is necessary to constrain the
average sum feedback rate. Let $B$ denote the number of bits sent
back by each feedback user and $K$ the number of feedback users. It
follows that the instantaneous sum feedback rate is $BK$. Since $B$
is a constant and $K$ a random variable, the constraint of the
average sum feedback rate can be written as
\begin{equation}
\text{(Sum Feedback Rate Constraint)}\quad BE[K] \leq R,
 \label{Eq:FBConst}
\end{equation}
where $R$ is the sum feedback rate constraint.

To simplify the analysis of the proposed algorithms in
Section~\ref{Section:Algo}, we make the following assumption about
the multi-user channels:
\begin{assumption}\label{AS:DLChan}\emph{
The downlink channel $\bh_u$ is an i.i.d. vector whose coefficients
are $\mathcal{CN}(0,1)$.}
\end{assumption}
Given this assumption, which is commonly made in the literature of
SDMA and multi-user diversity
\cite{SharifHassibi:CapMIMOBroadcastPartSideInfo:Feb:05,
ChoiForenza:OppSDMABeamSel:06,
Jindal:MIMOBroadcastFiniteRateFeedback:06,
SwannackWorenell:BroadcastChanLimitedCSI:2005,
YooGoldsmith:OptimBroadcastZeroForcingBeam:2006}, the channel
direction vector of each user follows a uniform distribution, which
greatly simplifies the design of feedback thresholds in
Section~\ref{Section:FbTh} and capacity analysis in
Section~\ref{Section:FbTh}.

\section{Algorithms}\label{Section:Algo}
OSDMA-TF is comprised of (i) CSI quantization at the subscribers,
(ii) selection of feedback users using feedback thresholds, and
(iii) joint beamforming  and scheduling at the base station. The
algorithms for performing these functions are discussed in
Section~\ref{Section:Quant} to Section~\ref{Section:BeamSel},
respectively. Furthermore, OSDMA-TF is compared with existing SDMA
algorithms in Section~\ref{Section:AlgoCmp}.

\subsection{CSI Quantization}
\label{Section:Quant} Without loss of generality, the discussion in
this section is focused on the $u$th user and the same algorithm for
CSI quantization is used by other users. For simplicity, we assume:
\begin{assumption}\label{AS:RXCSI}
\emph{The $u$th user has perfect receive CSI $\bh_u$.}
\end{assumption}
This assumption allows us to neglect channel estimation error at the
$u$th mobile. For convenience, the CSI, $\bh_u$, is decomposed into
two components: the \emph{gain} and the \emph{shape}, which are
quantized separately. Hence, $\bh_u = g_u\bs_u$ where $g_u =
\|\bh_u\|$ is the gain and $\bs_u = \bh_u/\|\bh_u\|$ is the shape.
The channel shape $\bs_u$ is quantized and sent back to the base
station for choosing beamforming vectors. The channel gain $g_u$ is
used for computing SINR, which is also quantized and sent back as a
channel quality indicator. Due to the ease of quantizing SINR that
is a scalar, we make the following assumption:
\begin{assumption}\label{AS:QuantErr}\emph{
The SINR is perfectly known to the base station through feedback.}
\end{assumption}
The same assumption is made in
\cite{SharifHassibi:CapMIMOBroadcastPartSideInfo:Feb:05,ChoiForenza:OppSDMABeamSel:06}.
This assumption allows us to focus our discussion on quantization of
the channel shape $\bs_u$.

Quantization of the channel shape $\bs_u$ is the process of matching
it to a member of a set of pre-determined vectors, called \emph{a
codebook}. Different from
\cite{Jindal:MIMOBroadcastFiniteRateFeedback:06,Huang:JointBeamScheduleLimtFb:06}
where code vectors are randomly generated, we propose a structured
codebook constructed as follows. The codebook, denoted as
$\mathcal{F}$, is comprised of $M$ sub-codebooks:
$\mathcal{F}=\cup_{m=1}^M\mathcal{F}_m$, each of which is comprised
of $N_t$ orthogonal vectors. The sub-codebooks
$\mathcal{F}_1,\mathcal{F}_2,\cdots,\mathcal{F}_M$ are independently
and randomly generated for example using the method in
\cite{Zyczkowski:RandUnitMatrices:94}. Each sub-codebook provides a
potential set of orthogonal beamforming vectors for downlink
transmission. Given a codebook $\mathcal{F}$ thus generated, the
quantized channel shape, denoted as $\uh_u$,  is the member of
$\mathcal{F}$ that forms the smallest angle with the channel shape
$\bs_u$ \cite{LovHeaETAL:GrasBeamMultMult:Oct:03}. Mathematically,
\begin{equation}\label{Eq:Quant}
    \uh_u =\mathcal{Q}(\bs_u)=
    \arg\max_{\bff\in\mathcal{F}}\left|\bff^\dagger\bs_u\right|,
\end{equation}
where the function $\mathcal{Q}$ represents the CSI quantization
process. We define the \emph{quantization error} as
\begin{equation}
    \text{(Quantization Error)} \quad \delta_u = \sin^2(\angle(\uh_u,
    \bs_u)\label{Eq:QuantErr:Def}).
\end{equation}
It is clear that the quantization error is zero if $\uh_u = \bs_u$.

\subsection{Feedback Algorithm}\label{Section:FBScheme}
To satisfy the sum feedback rate constraint \eqref{Eq:FBConst}, we
propose a threshold-based feedback algorithm, which allows only
users with good channels (high SINRs) to send back their CSI to the
base station. The SINR for the $u$th user is a function of the
channel power $\rho_u=\|\bh_u\|^2$ and the quantization error
$\delta_u$ in \eqref{Eq:QuantErr:Def} (see also
\cite{Huang:JointBeamScheduleLimtFb:06}):
\begin{equation}
    \SINR_u =
    \frac{1+P\rho_u}{1+P\rho_u\delta_u}-1. \label{Eq:SINR}
\end{equation}
Therefore, the feedback algorithm employs two feedback thresholds
for feedback user selection: the channel power threshold, denoted as
$\gamma$, and the quantization error threshold, denoted as
$\epsilon$. It follows that the $u$th user meets the feedback
criteria if $\rho_u \geq \gamma$ and $ \delta_u\leq \epsilon$. The
thresholds $\gamma$ and $\epsilon$ are designed in
Section~\ref{Section:FbTh} such that the sum feedback rate
constraint in \eqref{Eq:FBConst} is satisfied.

Given that the $u$th user meets the feedback thresholds, the
quantized channel shape $\uh_u$ is sent back to the base station
through a finite-rate feedback channel
\cite{LovHeaETAL:WhatValuLimiFee:Oct:2004,LovHeaETAL:GrasBeamMultMult:Oct:03}\footnote{The
feedback of SINR is ignored due to AS~\ref{AS:QuantErr}.}. Since the
quantization codebook $\mathcal{F}$ can be known \emph{a priori} to
both the base station and mobiles, only the index of $\uh_u$ needs
to be sent back. Therefore, the number of feedback bits per user is
$\log_2N$ since $|\mathcal{F}|=N$.

\subsection{Joint Beamforming and Scheduling}\label{Section:BeamSel}
Among feedback users, the base station schedules a subset of users
for downlink transmission using the criterion of maximizing sum
capacity and under the constraint of orthogonal beamforming. To
facilitate the description of the procedure for joint beamforming
and scheduling, we group feedback users according to their quantized
channel shapes by defining the following index sets:
\begin{equation}
    \mathcal{I}_{m,n} = \{1\leq u\leq U\mid \rho_u\geq \gamma, \ \delta_u\leq
    \epsilon, \ \mathcal{Q}(\bs_u)=\bff_{m,n}\}, \quad 1\leq m\leq M, \ 1\leq n\leq
    N_t,\label{Eq:IndexSet}
\end{equation}
where $\mathcal{Q}(\cdot)$ is the quantization function in
\eqref{Eq:Quant} and $\bff_{m,n}\in\mathcal{F}$ is the $n$th
member in the $m$th sub-codebook $\mathcal{F}_m\subset
\mathcal{F}$. The base-station adopts a two-step procedure for
joint beamforming and scheduling. First, it selects the user with
maximum SINR from each index set defined \eqref{Eq:IndexSet}.
Second, from these selected users, the base station schedules up
to $N_t$ users for downlink transmission under the constraint of
orthogonal beamforming. Following this procedure, the resultant
sum capacity can be written as
\begin{equation}
\mathcal{C} =  E\left[\max_{m=1,\cdots,M}\suml{n=1}{N_t}\log_2(1+
    \max_{u \in \mathcal{I}_{m,n}}\SINR_u)\right],\label{Eq:SumRateExp}
\end{equation}
where the two ``max" operators correspond to the two steps in the
procedure for joint beamforming and scheduling. In the event that
an index set $\mathcal{I}_{m,n}$ is empty, we set $\max_{u \in
\mathcal{I}_{m,n}}\SINR_u=0$ in \eqref{Eq:SumRateExp}.

\subsection{Comparison with Existing
Algorithms}\label{Section:AlgoCmp} We summarize in
Table~\ref{Tab:AlgoCmp} the key differences between OSDMA-TF and
existing SDMA algorithms with all-user feedback, including OSDMA-LF,
OSDMA-BS and OSDMA. Performance comparisons by Monte Carlo
simulation are provided in Section~\ref{Section:Numerical}.

\begin{threeparttable}[h]
  \centering
    \caption{Comparison of OSDMA-TF, OSDMA-LF, OSDMA-S and OSDMA}\label{Tab:Design}
  \begin{tabular}{r|cccc}
    & OSDMA-TF & OSDMA-LF & OSDMA-BS & OSDMA \\
    \hline
  \# of Feedback Users & $NN_t$ & $U$ & $U$  & $U$  \\
  Feeback/User (bits)~\tnote{a}&$2B + \log_2N$ &$2B + \log_2N$ & $B+I\log_2N_t$  & $B+\log_2N_t$  \\
  Sum Capacity (bits/s/Hz)~\tnote{b} & largest (7.5)& largest (7.5) & moderate (6.4)  & smallest
  (6.2)\\
  Beamforming \& Scheduling &centralized & centralized & distributed & N/A
  \end{tabular}
  \begin{tablenotes}\setlength{\baselineskip}{12pt}
\item[a] {\footnotesize Assume $B$ bits are required for
quantizing a channel gain and the quantization error of the channel
shape .} \item[b] {\footnotesize Sum capacity is computed for
$U=20$, $N_t=4$ and $\SNR=10dB$. Following
\cite{ChoiForenza:OppSDMABeamSel:06}, the sum capacity is reduced by
the feedback overhead factor $\alpha = 5\%$ for each round of CSI
feedback.} \item[c] {\footnotesize Refer to possibility that
different users select a same beamforming vector.}
\end{tablenotes}
\label{Tab:AlgoCmp}
\end{threeparttable}

\section{Feedback Design}\label{Section:FBDesign}
In Section~\ref{Section:FbTh}, the feedback thresholds for OSDMA-TF
(cf. Section~\ref{Section:FBScheme}) are designed as functions of
the number of users under the sum feedback constraint in
\eqref{Eq:FBConst}. Even if this constraint is satisfied, it is
likely that the instantaneous sum feedback rate exceeds the maximum
allowable feedback rate of the feedback channel and hence causes an
overflow. In Section~\ref{Section:OvFlow}, an upper-bound for the
feedback overflow probability is derived, which is useful for
designing the maximum feedback rate for the feedback channel.

\subsection{Feedback Thresholds}\label{Section:FbTh}
The feedback probability of each user is derived as a function of
the feedback thresholds. Subsequently, since the sum feedback rate
is proportional to this probability, we can thus derive the feedback
thresholds for a given sum feedback rate.

The feedback probability of a user is defined as the probability
that the user's channel power and quantization error meet the
respective thresholds. We focus on the feedback probability of a
single user since the channels of different users are i.i.d. given
AS~\ref{AS:DLChan} and hence the feedback probabilities of different
users are identical. For simplicity, we omit the user index, hence
the subscript $u$, for all notation in this section. Given
AS~\ref{AS:DLChan}, the channel power $\rho =\|\bh\|^2$ and the
channel direction $\bs = \bh/\|\bh\|$ are independent. It follows
that the two events,  namely the channel power and quantization
error thresholds are met, are independent. Therefore, we can derive
their probabilities separately. First, the probability that the
channel power $\rho$ of a user meets the power threshold is obtained
as
\begin{equation}
    P_{\gamma} = \Pr\{\rho\geq\gamma\}=\int_\gamma^\infty
    f_\rho(\rho)d\rho,\label{Eq:ProbGam}
\end{equation}
where $f_\rho(\rho)$ is the chi-squared PDF function given as
\begin{equation}\label{Eq:ChiPDF}
    f_{\rho} = \frac{\rho^{L-1}e^{-\rho}}{(L-1)!}.
\end{equation}
Second, the probability for meeting the quantization error
threshold, denotes as $P_\epsilon$, is obtained. To this end, we
define a set for each member of the codebook $\mathcal{F}$ as
\begin{equation}
    \mathcal{V}_n = \{\|\bv\|=1\mid 1-|\bv^H\bff_n|^2\leq \epsilon\}, \quad n =
    1,2,\cdots, N.\label{Eq:Cone}
\end{equation}
Intuitively, the set  $\mathcal{V}_n$ can be viewed as a ``cone"
with the radius $\epsilon$ and the axis $\bff_n$. The probability,
$P_\epsilon$, can be defined in terms of these sets as
\begin{equation}
    P_\epsilon = \Pr\{\bs \in \cup_n \mathcal{V}_n\}.
\end{equation}
By applying the union bound and using the symmetry of different
users' channels,
\begin{equation}
    P_\epsilon \leq \suml{n=1}{N}\Pr\{\bs \in  \mathcal{V}_n\}= N\Pr\{\bs \in  \mathcal{V}_n\}.
\end{equation}
By denoting $1-|\bs^H\bff_n|^2$ as $\delta$ and from
\eqref{Eq:Cone},
\begin{equation}
P_\epsilon \leq N\Pr\{\delta \leq \epsilon\} =N\epsilon^{L-1},
\label{Eq:PrDirBnd}
\end{equation}
where we use the following result from
\cite{Jindal:MIMOBroadcastFiniteRateFeedback:06},
\begin{equation}
    \Pr\{\delta \leq \epsilon\}=\epsilon^{L-1}\label{Eq:QuantErr}.
\end{equation}
By combining \eqref{Eq:ProbGam} and \eqref{Eq:PrDirBnd}, the
feedback probability for each user is given in the following lemma.
\begin{lemma}\label{Lem:FBPProb}
The feedback probability for each user is given as
\begin{equation}\label{Eq:FBProb}
    P_v = P_\gamma P_\epsilon \leq N\epsilon^{L-1}\int_\gamma^\infty
    f_\rho(\rho)d\rho.
\end{equation}
\end{lemma}

The sum feedback rate, denoted as $R$, can be expressed as $R =
E[K]B$ where $E[K]$ denotes the average number of feedback users and
$B$ the number of bits sent back by each of them. Furthermore,
$E[K]$ can be written as
\begin{equation}
    E[K] = UP_v,\label{Eq:FBUsr}
\end{equation}
with $P_v$ given in \eqref{Eq:FBProb}.
 With $B$ fixed, the sum feedback rate is proportional to $E[K]$.
We derive a set of feedback thresholds such that $E[K]$ is limited
by an upper-bound, which  is independent of the number of users
$U$. By choosing a proper value for the upper-bound, we can thus
satisfy any given constraint on the sum feedback rate $R$. These
results are shown as the following theorem.

\begin{theorem}\label{Theo:FBRate}\emph{
Consider the following channel power and quantization error
thresholds
\begin{eqnarray}
    \gamma &=& \log U - \lambda\log\log U, \quad \lambda > 0,
    \label{Eq:GamTh}\\
    \epsilon &=& \left[U^{1-\phi}(\log U)^{\phi\lambda}\right]^{-1/(L-1)}\label{Eq:EpsTh},
\end{eqnarray}
where
\begin{equation}
            \phi = -\gamma\ln\left(\frac{1}{L!}\int_\gamma^\infty
        \rho^{L-1}e^{-\rho}d\rho\right).\label{Eq:GamFunExp}
\end{equation}
Given these thresholds,   the average number of feedback users
$E[K]$ is upper-bounded as
\begin{equation}
E[K] \leq  NN_t,
\end{equation}
where $N$ is the cardinality of the CSI quantization codebook
$\mathcal{F}$. }
\end{theorem}
\begin{proof}
The theorem follows by substitution of the feedback thresholds in
\eqref{Eq:GamTh} and \eqref{Eq:EpsTh} into \eqref{Eq:FBProb} and
then \eqref{Eq:FBUsr}.
\end{proof}
A few remarks are in order:
\begin{itemize}
\item Given the feedback thresholds in Theorem~\ref{Theo:FBRate},
the sum feedback rate is bounded as
\begin{equation}\label{Eq:FBRate}
    R\leq BNN_t,
\end{equation}
where $B=B_s + \log_2N$ is the number of feedback bits per user with
$B_s$ is the number of bits for quantizing the SINR
feedback\footnote{Usually, $B_s << \log_2N$ since SINR is a scalar
while the channel direction is a vector.} (cf.
Section~\ref{Section:Algo}).

\item The power and quantization thresholds in \eqref{Eq:GamTh} and
\eqref{Eq:EpsTh} are chosen jointly to ensure the capacity of each
scheduled user grows with the number of users $U$ at an optimal
rate, namely $\log_2\log_2U$
\cite{SharifHassibi:CapMIMOBroadcastPartSideInfo:Feb:05}. Detailed
analysis is given in Section~\ref{Section:CapAna}.

\item The parameter $\lambda$ in \eqref{Eq:GamTh} and
\eqref{Eq:EpsTh} affects the signal-to-interference ratio (SIR) of
feedback users. Its optimal value for maximizing sum capacity can
be chosen via numerical methods since analytical methods seem
difficult.

\item CSI quantization error causes interference between
simultaneous users and can potentially prevent the sum capacity from
increasing with the number of users as observed in
\cite{DingLove:SubspaceFbBroadcastChan:2005,Jindal:MIMOBroadcastFiniteRateFeedback:06}.
This motivates the design of the quantization error threshold in
\eqref{Eq:EpsTh}. This threshold ensures that the quantization error
of each feedback user converges to zero with the number of users
$U$. This result is proved shortly.
\end{itemize}

To prove that the quantization errors of feedback users diminishes
with the number of users $U$, we require the Alzer's bounds for the
gamma function  rewritten as the following lemma
\cite{Alzer:GamFunIneq:97}
\begin{lemma}[Alzer's Inequality]\label{Lem:Alzer}\emph{
The incomplete Gamma function is bounded as
\begin{equation}\label{Eq:Alzer}
    \left[1-e^{-\beta \gamma}\right]^{N_t} <
    \int_0^\gamma f_\rho(\rho)d\rho < \left[1-e^{-\gamma}\right]^{N_t},
\end{equation}
where $\beta = ({N_t}!)^{-1/{N_t}}$ and $f_{\rho}(\rho)$ is the
chi-squared PDF in \eqref{Eq:ChiPDF}.}
\end{lemma}
Using this lemma, we obtain the result as shown in the following
corollary of Theorem~\ref{Theo:FBRate}.
\begin{corollary}\label{Cor:PrGam}\label{Cor:QuantErr}\emph{
The quantization error threshold $\epsilon$ in \eqref{Eq:EpsTh}
ensures the quantization error of a feedback user, $\delta$,
converges to zero with the number of users $U$
\begin{equation}
    \lim_{U\rightarrow\infty}\delta \leq
    \lim_{U\rightarrow\infty}\epsilon = 0.
\end{equation}
}
\end{corollary}
\begin{proof}
See Appendix~\ref{APP:QuantErr}.
\end{proof}

Last, we provide the following proposition, which shows that the
upper-bound on the average number of feedback users $E[K]$ is
tight if the number of users $U$ is large. We define the minimum
distance between any two members of the codebook $\mathcal{F}$ as
\begin{equation}\label{Eq:MinDist}
    \Delta\delta_{\min} = \min_{1\leq a, b\leq
    N}[1-|\bff_a^H\bff_b|^2],
\end{equation}
where $\bff_a \in \mathcal{F}$ and $\bff_b\in\mathcal{F}$.
\begin{proposition}\label{Prop:FBUsr}\emph{
For any codebook $\mathcal{F}$ with $\Delta\delta_{\min}>0$, there
exists an integer $U_0$ such that $\forall \ U \geq U_0$, the
average number of feedback users $E[K]$ is given as
\begin{equation}\label{Eq:K}
    E[K] = NN_t.
\end{equation}}
\end{proposition}
\begin{proof}
See Appendix~\ref{APP:FBUsr}.
\end{proof}

\begin{figure}[h]
\centering
  \includegraphics[width=11cm]{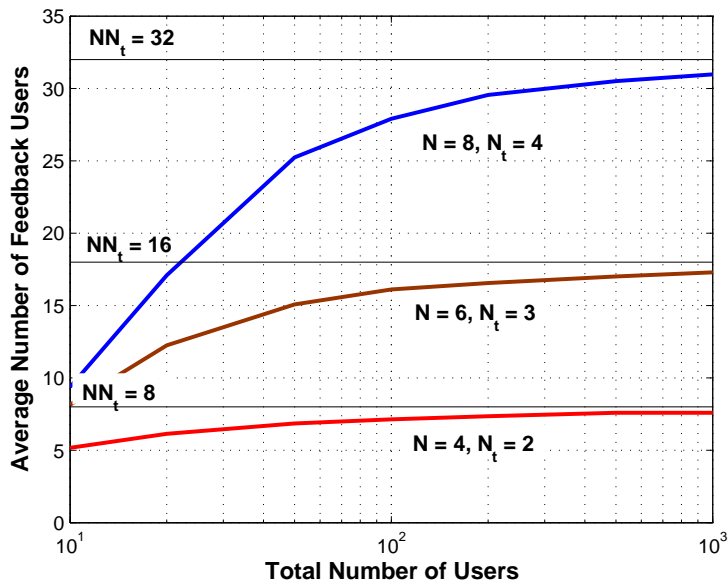}\\
  \caption{Average numbers of feedback users for OSDMA-TF}\label{Fig:FbUsr}
\end{figure}

For illustration of Theorem~\ref{Theo:FBRate} and
Proposition~\ref{Prop:FBUsr}, the numbers of feedback users $E[K]$
averaged over different channel realizations and randomly
generated codebooks are plotted against different numbers of users
$U$ in Fig.~\ref{Fig:FbUsr}. First, $E[K]$ is observed to be
upper-bounded by $NN_t$, which agrees with
Theorem~\ref{Theo:FBRate}. Second, $E[K]$ converges to $NN_t$ with
the number of users $U$, which verifies
Proposition~\ref{Prop:FBUsr}.

\subsection{Overflow Probability for Feedback Channel }\label{Section:OvFlow}
The overflow probability is defined as the probability that the
instantaneous sum feedback rate exceeds the average sum feedback
rate. A small overflow probability reduces the average waiting time
of feedback users and improves the system stability
\cite{BertsekasBook:DataNetwk:92}. In this section, we show that an
arbitrarily small overflow probability can be maintained by making
the maximum allowable feedback rate sufficiently large relative to
the average sum feedback rate.

The multiuser feedback channel satisfying the sum feedback rate
constraint can be implemented \emph{asynchronously} or
\emph{synchronously}. For the first case, the contention feedback
method as in
\cite{TangHeath:OppFbDLMuDiv:2005,TangHeath:OppFbMuMIMOLinearRx:2005}
can be applied, which allows feedback users to compete for uplink
transmission. For the second case, multiuser feedback is coordinated
by a base station following a multiple-access scheme, such as
\emph{orthogonal frequency division multiple access} (OFDMA),
\emph{time division multiple access} (TDMA), or \emph{code division
multiple access} (CDMA) \cite{RappaportBook:WirelssComm:01}. For
both a synchronous and an asynchronous feedback channel, a small
feedback overflow probability is desirable for the reasons stated
earlier.

For simplicity and due to their equivalence, we measure the
\emph{instantaneous, average} and  \emph{maximum allowable} sum
feedback rate using the instantaneous, average and  maximum
allowable numbers of feedback users,  denoted as $K$, $E[K]$ and
$K_{\max}$, respectively. The overflow probability can be
upper-bounded using the Chernoff bound
\cite{Janson:RandomGraph:Book} as follows. For each user, we define
a Bernoulli random variable $T_u$ indicating whether the user meets
the feedback thresholds
\begin{equation}\label{Eq:Bernoulli}
    T_u = 1\{\delta_u\leq\epsilon \ \text{and} \ \rho_n \geq
    \gamma\}, \quad u = 1,2,\cdots, U.
\end{equation}
The instantaneous number of feedback user, $K$, can be expressed as
the sum of  these Bernoulli random variables, hence $ K=
\sum_{u=1}^{U}T_u$. Using the Chernoff bound for the summation of
i.i.d. Bernoulli random variables derived in
\cite{Janson:RandomGraph:Book}, we can obtain an upper-bound for the
overflow probability.

\begin{proposition}\label{Prop:Overflow}\emph{
The overflow probability of the feedback channel is upper-bounded as
\begin{equation}
\Pr(K\geq K_{\max}) \leq
\exp\left[-K_{\max}\log\left(\frac{K_{\max}}{E[K]}\right)-(U-K_{\max})
\log\left(\frac{U-K_{\max}}{U-E[K]}\right)\right],\label{Eq:CherBnd}
\end{equation}
where $K_{\max}$ is the maximum number of feedback users supported
by the feedback channel and $E[K] \leq NN_t$.}
\end{proposition}
The upper-bound obtained above is useful for determining the maximum
data rate the feedback channel should support such that a constraint
on the overflow probability is satisfied since the feedback data
rate is proportional to the number of feedback users.

\begin{figure}[h]
\centering \subfigure[${N_t} = 2, N = 4$]{
\includegraphics[width=8cm]{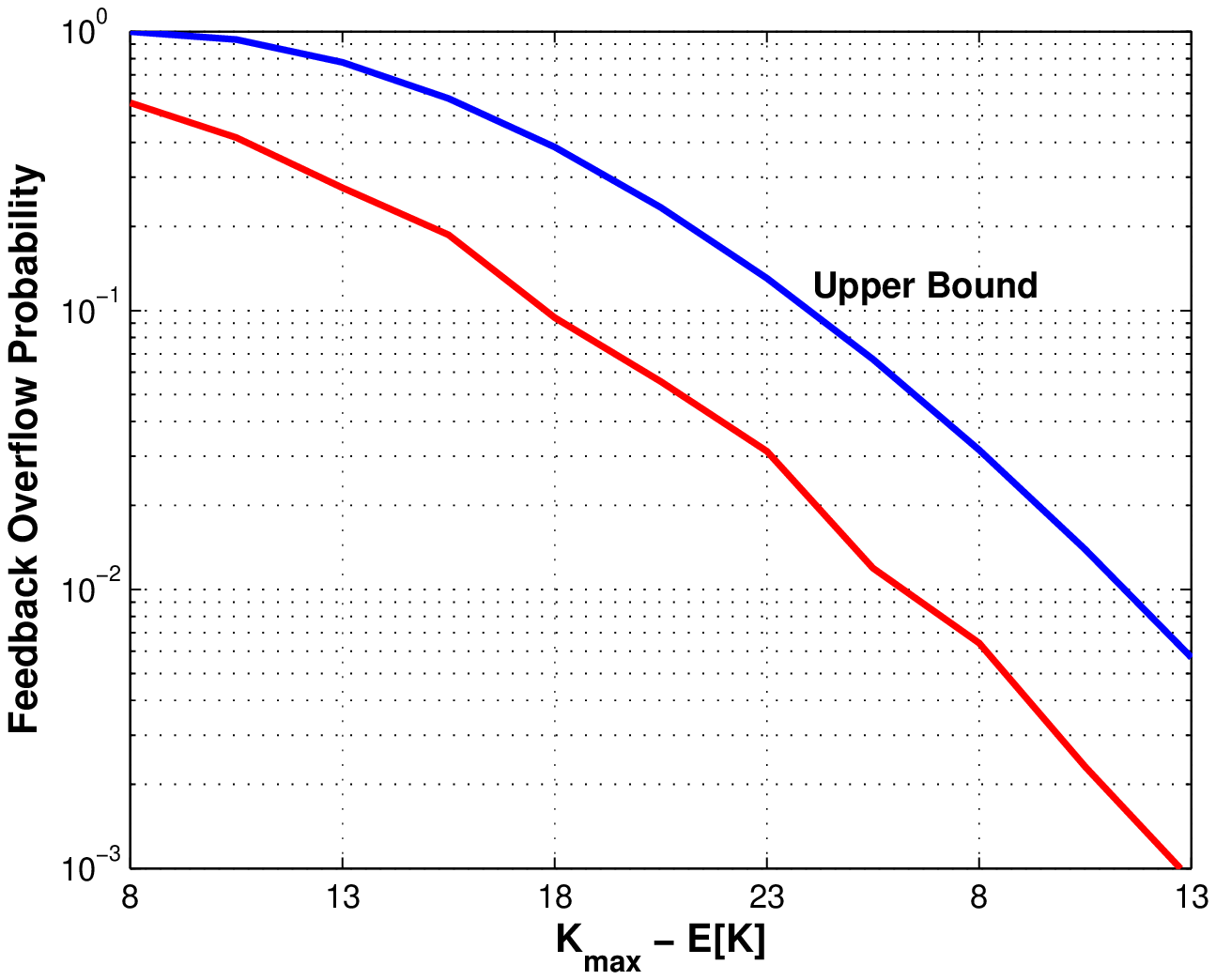}}
\subfigure[${N_t} = 2, N = 2$]{
\includegraphics[width=8cm]{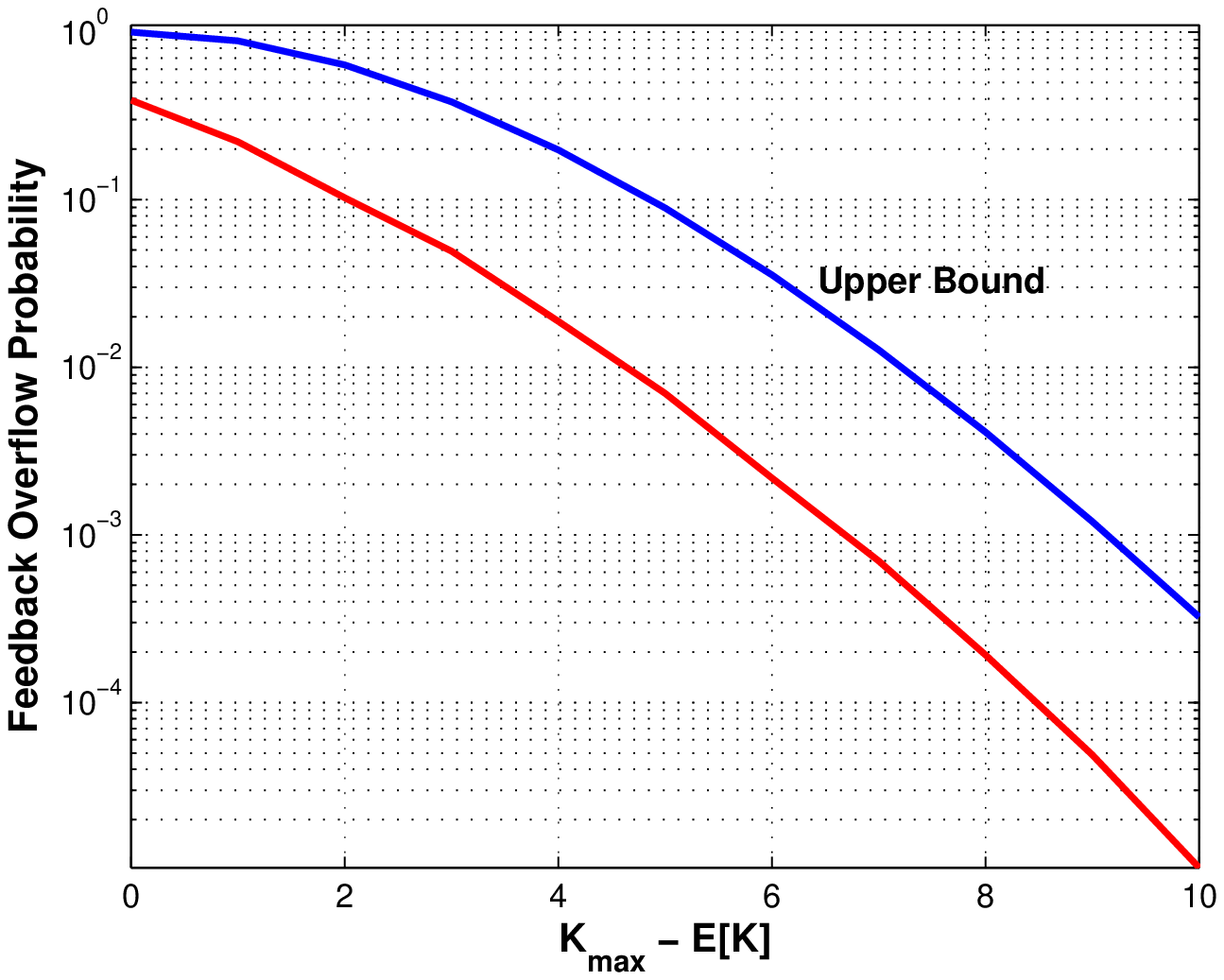}}
  \caption{Feedback channel overflow probability for OSDMA-TF}\label{Fig:ProbOverFlow}
\end{figure}

 In Fig.~\ref{Fig:ProbOverFlow}, the upper bound in \eqref{Eq:CherBnd} is
compared with the actual overflow probability obtained by Monte
Carlo simulation. It can be observed that both the overflow
probability and its upper-bound decreases at the same slope and
approximately exponentially with the difference $(K_{\max}-E[K])$.

\section{Analysis of Sum Capacity}\label{Section:CapAna}
In this section, for a large number of users ($U\rightarrow\infty$),
we show that the sum capacity of OSDMA-TF can grow at a rate close
to the optimal one, namely $N_t\log_2\log_2N$, if the sum feedback
rate is sufficiently large.

Before proving the main result, several useful lemmas are
presented. As we know, the high sum capacity of SDMA is due to its
ability of supporting up to $N_t$ simultaneous users. The first
lemma concerns the probability for the impossibility of scheduling
$N_t$ users. This probability is name \emph{probability of
scheduled user shortage} and denoted as $P_{\beta}$. Using the
index sets defined in \eqref{Eq:IndexSet}, we can express
$P_{\beta}$ as
\begin{equation}
    P_\beta = \Pr\left\{\max_{1\leq m\leq
M}\suml{n=1}{N_t}1\{\mathcal{I}_{m,n}\neq \emptyset \}<
N_t\right\}.\label{Eq:PrUsrShort:Def}
\end{equation}
It can be upper-bounded as shown in Lemma~\ref{Lem:PrUsrShort}.
\begin{lemma}\label{Lem:PrUsrShort}\emph{
 The probability of scheduled user shortage is upper-bounded as
\begin{equation}\label{Eq:PrNoFb}
P_\beta  < (N_t e^{-N_t})^M.
\end{equation}}
\end{lemma}
\begin{proof}
See Appendix~\ref{APP:PrUsrShort}.
\end{proof}
As to be shown later, the probability $P_\beta$ characterizes the
decrease of the asymptotic growth rate of the sum capacity caused by
scheduled user shortage or equivalently the sum feedback rate
constraint.

From \eqref{Eq:SINR}, the multi-user interference encountered by a
scheduled user with channel power $\rho$ and channel quantization
error $\delta$ is $P\rho\delta$. Lemma~\ref{Lem:MeanProd} shows that
the average of the multi-user interference converges to zero with
the number of user $U$.
\begin{lemma} \label{Lem:MeanProd}\emph{Let $\rho$ and $\delta$ denote the channel power and
quantization error of a feedback user. We have
\begin{equation}
    \lim_{U\rightarrow\infty} E[\rho\delta \mid \rho \geq \gamma, \delta \leq \epsilon] = 0, \quad \text{if} \ \lambda\geq
    N_t-1,
\end{equation}
where $\lambda$ is the parameter of the power threshold in
\eqref{Eq:GamTh}. }
\end{lemma}
\begin{proof}
See Appendix~\ref{APP:MeanProd}.
\end{proof}

\begin{figure}
\centering
  \includegraphics[width=11cm]{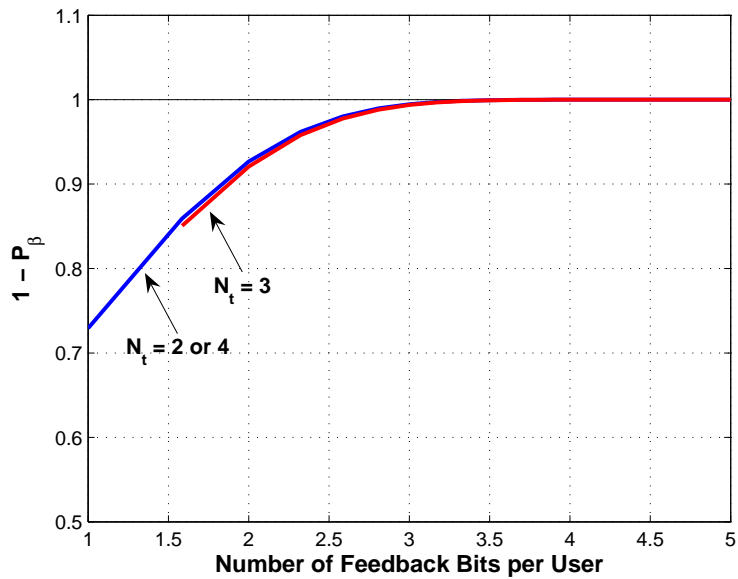}\\
  \caption{Convergence of the lower bound in \eqref{Eq:SumRate}, $1-P_\beta$, to one with the number of feedback bits
  per user $\log_2N$. }\label{Fig:SumCapRat}
\end{figure}

The main result of this section is the following theorem.
\begin{theorem}\label{Theo:CapOrder} \emph{For a large number of users
($U\rightarrow\infty$), the sum capacity of OSDMA-TF grows with the
number of transmit antennas $N_t$ linearly and with the number of
users double logarithmically
\begin{equation}\label{Eq:SumRate}
    1\geq \lim_{U\rightarrow\infty}\frac{\mathcal{C}}{N_t\log_2\log_2U} > 1 -
    (N_t e^{-N_t})^M, \quad \text{if} \ \lambda\geq N_t -1,
\end{equation}
where $\lambda$ is the parameter of the power threshold in
\eqref{Eq:GamTh}.}
\end{theorem}
\begin{proof}
See Appendix~\ref{APP:CapOrder}.
\end{proof}
The above theorem shows the effect of a sum feedback rate constraint
is to decrease the growth rate of the sum capacity with respect to
that for feedback from all users, namely $N_t\log_2\log_2U$
\cite{SharifHassibi:CapMIMOBroadcastPartSideInfo:Feb:05,Huang:JointBeamScheduleLimtFb:06}.
Nevertheless, such difference in growth rate can be made arbitrarily
small by increasing the sum feedback rate, or equivalently the
number of feedback bits per feedback user, as stated in the
following corollary.
\begin{corollary}\emph{
By increasing the number of feedback bits per feedback user
($\log_2N$), the sum capacity of OSDMA-TF can grow at the optimal
rate:
\begin{equation}
    \lim_{N\rightarrow\infty}\lim_{U\rightarrow\infty}\frac{\mathcal{C}}{N_t\log_2\log_2U} = 1, \quad \text{if} \ \lambda\geq N_t-1.
\end{equation}}
\end{corollary}
\begin{proof}
The result follows from \eqref{Eq:SumRate} and $N=MN_t$.
\end{proof}
We can observe from Fig.~\ref{Fig:SumCapRat} that the asymptotic
growth rate for the sum capacity for OSDMA-TF converges to the
optimal value, hence $1-P_\beta\rightarrow 1$ from
\eqref{Eq:SumRate}, very rapidly as the number of feedback bits per
user ($\log_2N$) increases.



%

\section{Performance Comparison}\label{Section:Numerical}
In this section, we compare the sum capacity and the sum feedback
rate of OSDMA-TF with the case of all-user feedback, which is
equivalent to OSDMA-TF with trivial feedback thresholds $\gamma = 0$
and $\epsilon = 1$. Similar comparisons are also conducted  between
OSDMA-TF and existing algorithms including OSDMA-LF
\cite{Huang:JointBeamScheduleLimtFb:06}, OSDMA-BS
\cite{ChoiForenza:OppSDMABeamSel:06} and OSDMA
\cite{SharifHassibi:CapMIMOBroadcastPartSideInfo:Feb:05}.

\begin{figure}
\centering \subfigure[Sum
Capacity]{\includegraphics[width=11cm]{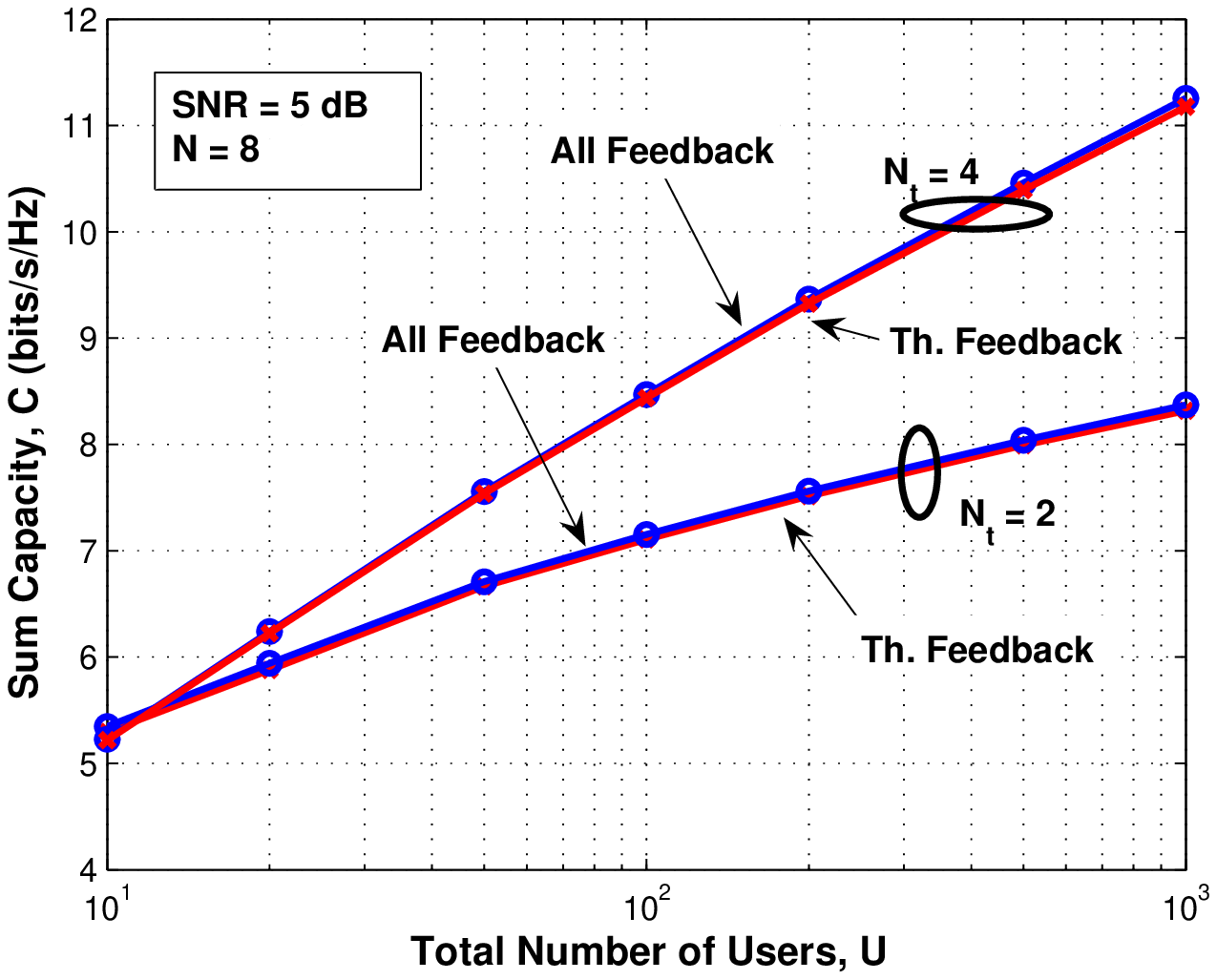}}\\
\subfigure[Sum Feedback
Rate]{\includegraphics[width=11cm]{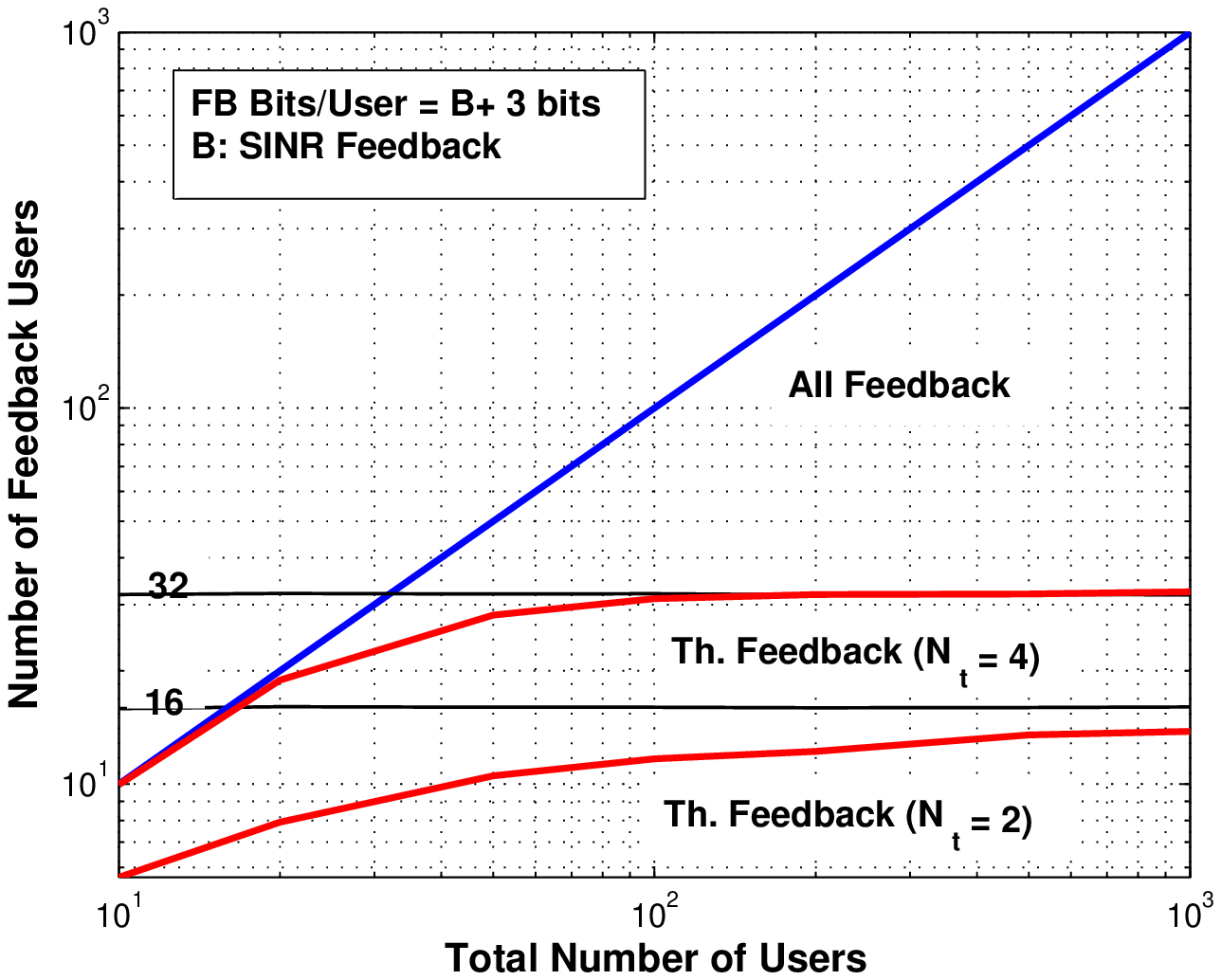}}\
  \caption{Sum capacities of threshold feedback (OSDMA-TF) and all-user feedback}\label{Fig:SumCap}
\end{figure}

The sum capacities of OSDMA-TF and the corresponding case of
all-user feedback are plotted against the number of users $U$ in
Fig.~\ref{Fig:SumCap}(a) for the cases of $N_t=\{2,4\}$ transmit
antennas. For these two cases, the parameter $\lambda$ for the power
threshold $\gamma$ in \eqref{Eq:GamTh} is assigned the values of 1
and 1.5 respectively, which are found numerically to be sum capacity
maximizing. Each user quantizes his/her channel shape using a
codebook of size $N = 8$ and hence each feedback user sends back
$\log_2N=3$ bits. It can be observed from Fig.~\ref{Fig:SumCap}(a)
that the sum feedback rate constraint for SDMA-TF incurs negligible
loss in sum capacity with respect to the case of all-user feedback.
Next, the number of feedback users for OSDMA-TF and all-user
feedback are compared in Fig.~\ref{Fig:SumCap}(b). Note that the sum
feedback rate $R$ is proportional to the number of feedback users
$E[K]$: $R=3E[K]$ bits. It can be observed that the number of
feedback users for OSDMA-TF is upper bounded by 32 for $N_t=4$ and
$16$ for $N_t=2$ since OSDMA-TF is designed for satisfying a sum
feedback constraint (cf. Section~\ref{Section:FbTh}). In summary,
OSDMA-TF achieves almost identical sum capacity as the case of all
user feedback but with a dramatic reduction on sum feedback rate for
a large number of users $U$.

\begin{figure}
\centering
\subfigure[Sum Capacity]{  \includegraphics[width=11cm]{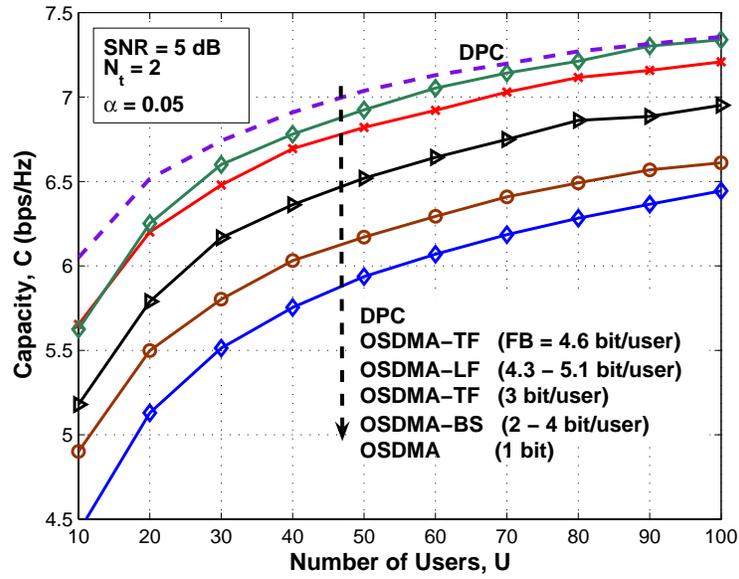}}\\
\subfigure[Sum Feedback Rate]{
\includegraphics[width=11cm]{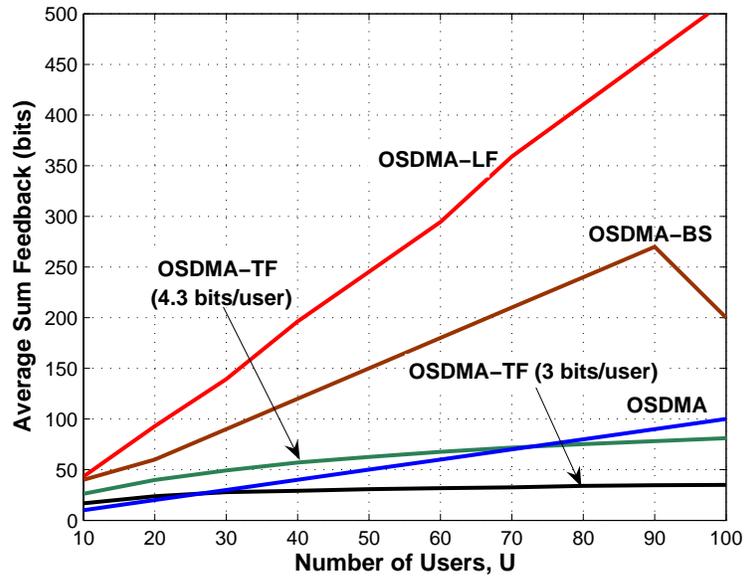}}\\
  \caption{Sum capacities and sum feedback rates of OSDMA-TF, OSDMA-LF, OSDMA-BS and OSDMA. }\label{Fig:SumCapCmp}
\end{figure}

In Fig.~\ref{Fig:SumCapCmp}, the sum capacity and sum feedback rate
of OSDMA-TF is compared with those of OSDMA-LF, OSDMA-BS and OSDMA
for different numbers of users $U$, with $N_t=2$ and an SNR of 5 dB.
The number of feedback bits per feedback user differs for the
algorithms in comparison since they use different sizes for
quantization codebooks or different feedback algorithms. For
OSDMA-TF, two codebook sizes $N=8$ and $N=24$ are considered,
corresponding to 3 and 4.6 feedback bits for each feedback user,
respectively\footnote{The average number of feedback users $\bar{K}$
is a function of $N$ (cf. Section~\ref{Section:FbTh}).}. For
OSDMA-LF, the codebook size $N$ increases with $U$ as:
$N=5\lceil(\log_2U)^{N_t-1}\rceil$ to avoid saturation of sum
capacity due to limited feedback
\cite{Huang:JointBeamScheduleLimtFb:06}. The codebook sizes for
OSDMA-BS and OSDMA are both $N=2$. Different from other algorithms,
the CSI feedback for OSDMA-BS is performed iteratively, where each
iteration penalizes  the sum capacity by a factor of
$0\leq\alpha\leq 1$ \cite{ChoiForenza:OppSDMABeamSel:06}. Let $I$
denote the number of feedback iterations. Therefore, for OSDMA-BS,
the total feedback for each user is $I$ bits and the sum capacity
with feedback penalty is given as $\mathcal{C}_p =
(1-I\alpha)\mathcal{C}$. For fair comparison, we also apply this
feedback penalty to the other algorithms\footnote{$I=1$ for
OSDMA-TF, OSDMA-LF and OSDMA}.

From Fig.~\ref{Fig:SumCapCmp}(a),  we can observe that OSDMA-TF
($N=24$) yields the highest sum capacity and OSDMA-TF ($N=8$) is
outperformed only by OSDMA-LF. Moreover, the sum capacity of
OSDMA-TF converges to DPC rapidly. From Fig.~\ref{Fig:SumCapCmp}(b),
the sum feedback rates for OSDMA-TF is observed to grow much more
gradually with the number of users $U$ than those for other
algorithms. Asymptotically, the sum feedback rates for OSDMA-TF
saturate due to sum feedback rate constraint (cf.
Theorem~\ref{Theo:FBRate}) while those for other algorithms continue
to increase with $U$. Several other observations can be made from
Fig.~\ref{Fig:SumCapCmp}(b). First, for $U\geq 25$, OSDMA-TF with
$N=8$ has the smallest sum feedback rate among all algorithms but it
outperforms OSDMA-BS and OSDMA in terms of sum capacity. Second, for
$U\geq 75$, OSDMA-TF with $N=24$ yields the highest sum capacity
among all algorithms and also requires the smallest sum feedback
rate. Third, the sum feedback rate for OSDMA-LF is the highest among
all algorithms. Fourth, the drop of sum feedback rate for OSDMA-BS
at $U=90$ is due to the decrease of the optimal number of feedback
iterations, which is obtained in
\cite{ChoiForenza:OppSDMABeamSel:06}.

\section{Conclusion}\label{Section:Conclusion}
This paper proposes a SDMA downlink algorithm with a sum feedback
rate constraint, which is applied by using feedback thresholds on
users' channel power and channel quantization errors. We derived the
expressions for these thresholds and the upper bound for the
corresponding feedback overflow probability. Furthermore, we
obtained the asymptotic growth rate of the sum capacity with the
number of users. We showed that it can be made arbitrarily close to
the optimal value by increasing the sum feedback rate. From
numerical results, we found that limiting the sum feedback rate
incurs negligible loss on sum capacity. Moreover, we demonstrated
that the proposed SDMA algorithm is capable of outperforming
existing algorithms despite having a much smaller sum feedback rate.

\appendix
\subsection{Proof of
Corollary~\ref{Cor:QuantErr}}\label{APP:QuantErr} From
Lemma~\ref{Lem:Alzer} and \eqref{Eq:GamFunExp}, we have
\begin{equation}
    \left(1-e^{-\beta\gamma}\right)^{N_t} < 1- {N_t}e^{-\phi\gamma} <
    \left(1-e^{-\gamma}\right)^{N_t}.\label{Eq:QErr:a}
\end{equation}
From the definition in \eqref{Eq:GamTh}, we observe that $\gamma$
monotonically increases with $U$ when $U$ is large. Therefore, from
\eqref{Eq:GamTh} and \eqref{Eq:QErr:a},
 there exists an integer $U_0$ such that $\forall \ U \geq U_0$,
\begin{equation}
    1-{N_t}e^{-\beta\gamma} < 1- {N_t}e^{-\phi\gamma} <
    1-{N_t}e^{-\gamma}.
\end{equation}
It follows that $\beta < \phi < 1$. Combining this with
\eqref{Eq:EpsTh}, the result in Corollary~\ref{Cor:QuantErr}
follows.

\subsection{Proof of Proposition~\ref{Prop:FBUsr}}\label{APP:FBUsr}
We observe from \eqref{Eq:FBUsr} and \eqref{Eq:FBProb} that
\begin{equation}
E[K] = UP_{\gamma}P_{\epsilon} \leq UNP_{\gamma}\Pr\{\bs \in
\mathcal{V}_n\},
\end{equation}
where $1\leq n \leq N$ is arbitrary. We will prove the existence of
an integer $U_0$ such that the equality in the above equation holds
$\forall \ U \geq U_0$. Following \eqref{Eq:EpsTh}, such an integer
exists such that
\begin{equation}
    \epsilon \leq \frac{\Delta\delta_{\min}}{2}, \quad \forall \
    U\geq U_0, \label{Eq:EpsMinDist}
\end{equation}
where $\Delta\delta_{\min}>0$ is the minimum distance for the
codebook $\mathcal{F}$ as defined in \eqref{Eq:MinDist}. Assume that
there exist two overlapping sets $\mathcal{V}_a$ and
$\mathcal{V}_b$, $\mathcal{V}_a \cap\mathcal{V}_b\neq \emptyset$.
Let $\bs \in \mathcal{V}_a \cap\mathcal{V}_b$. From the triangular
inequality and the definition in \eqref{Eq:MinDist},
\begin{equation}
    (1-|\bs^H\bff_a|^2)    + (1-|\bs^H\bff_b|^2) \geq
    \Delta\delta_{\min}. \label{Eq:Ineq1}
\end{equation}
On the other hand, from the definition in \eqref{Eq:Cone},
\begin{equation}
    (1-|\bs^H\bff_a|^2)    + (1-|\bs^H\bff_b|^2) \leq
    2\epsilon.\label{Eq:Ineq2}
\end{equation}
Nevertheless, \eqref{Eq:EpsMinDist} leads to the contradiction
between \eqref{Eq:Ineq1} and \eqref{Eq:Ineq2}. Thereby, we prove
that given \eqref{Eq:EpsMinDist},
\begin{equation}
\mathcal{V}_a \cap\mathcal{V}_b= \emptyset,\quad \forall \ U\geq U_0
\ \text{and} \ 1\leq a, b\leq N.
\end{equation}
It follows that
\begin{equation}
    \Pr\{\bs \in \cup_{n}\mathcal{V}_n\} = N\Pr\{\bs \in
    \mathcal{V}_1\}.\label{Eq:UniBnd}
\end{equation}
Therefore,
\begin{equation}
    K = UNP_{\gamma}\Pr\{\bs \in \mathcal{V}_1\}, \quad \forall \ U\geq U_0.
\end{equation}
Substitution of \eqref{Eq:GamTh} and \eqref{Eq:EpsTh} into the above
equation completes the proof.

\subsection{Proof of
Lemma~\ref{Lem:PrUsrShort}}\label{APP:PrUsrShort} From
\eqref{Eq:PrUsrShort:Def},
\begin{eqnarray}
P_\beta &=&
\Pr\left\{\bigcap_{m=1}^M\left\{\suml{n=1}{N_t}1\{\mathcal{I}_{m,n}\neq
\emptyset\}< N_t\right\}\right\},\label{Eq:PfPrBet:a}\\
&\overset{(a)}{=}&\prod_{m=1}^M\Pr\left\{\suml{n=1}{N_t}1\{\mathcal{I}_{m,n}\neq
\emptyset\}< N_t\right\},\nn\\
&=&\prod_{m=1}^M\Pr\left\{\bigcup_{n=1}^{N_t}\left\{\mathcal{I}_{m,n}=
\emptyset\right\}\right\},\nn\\
&\overset{(b)}{\leq}&\prod_{m=1}^M\left(N_t\Pr\left\{\mathcal{I}_{m,n}=
\emptyset\right\}\right),\nn\\
&\overset{(c)}{=}&\prod_{m=1}^M\left[N_t(1-P_{\gamma}\Pr\{\delta\leq \epsilon\})^U\right],\nn\\
&\overset{(d)}{=}&\left[N_t(1-N_t/U)^U\right]^M,\nn\\
&\leq&\left[N_te^{-N_t}\right]^M.\nn
\end{eqnarray}
The equality (a) results from the independence of the $M$ events in
\eqref{Eq:PfPrBet:a} due to the independent generations of the $M$
sub-codebook in the codebook $\mathcal{F}$. The inequality (b) is
obtained by applying the union bound as well as using the equal
probabilities of the events $\left\{\mathcal{I}_{m,n}=
\emptyset\right\}$ for $m=1,2,\cdots, M$. The equality (c) follows
from the definition of the set $\mathcal{I}_{m,n}$ in
\eqref{Eq:IndexSet}. The equality (d) is obtained from
\eqref{Eq:ProbGam}, \eqref{Eq:QuantErr}, \eqref{Eq:GamTh} and
\eqref{Eq:EpsTh}.

\subsection{Proof of Lemma~\ref{Lem:MeanProd}}\label{APP:MeanProd}
Given AS~\ref{AS:DLChan}, $\rho$ and $\delta$ are independent, hence
$E[\rho\delta \mid \rho \geq \gamma, \delta \leq \epsilon] = E[\rho
\mid \rho \geq \gamma]E[\delta \mid \delta \leq \epsilon]$. By
definition,
\begin{equation}
E[\rho\mid \rho \geq \gamma] = \frac{\int_\gamma^\infty \rho\cdot
\rho^{{N_t}-1}e^{-\rho}d\rho}{\int_\gamma^\infty
\rho^{{N_t}-1}e^{-\rho}d\rho} = \frac{\Gamma({N_t}+1,
\gamma)}{\Gamma({N_t}, \gamma)},\nn
\end{equation}
where $\Gamma(\cdot, \cdot)$ denotes the incomplete Gamma function
\cite{Alzer:GamFunIneq:97}. By expanding $\Gamma(\cdot, \cdot)$, we
obtain an upper-bound for $E[\rho\mid \rho \geq \gamma]$ as
\begin{eqnarray}
E[\rho\mid \rho \geq \gamma] &=&
\frac{{N_t}!e^{-\gamma}\sum_{i=0}^{N_t}\gamma^i/i!}{({N_t}-1)!e^{-\gamma}\sum_{i=0}^{{N_t}-1}\gamma^i/i!},\nn
\\
&=& {N_t}\left(1 +
\frac{\gamma^{N_t}/{N_t}!}{\sum_{i=0}^{{N_t}-1}\gamma^i/i!}\right),\nn\\
&<& {N_t}\left(1 + \frac{\gamma^{N_t}/{N_t}!}{\gamma^{{N_t}-1}/({N_t}-1)!}\right),\nn\\
&=& {N_t} + \gamma. \label{Eq:PwrMean}
\end{eqnarray}
Next, we obtain the expression of $E[\delta \mid \delta \leq
\epsilon]$ as:
\begin{eqnarray}
E[\delta\mid \delta \leq \epsilon] & = & \int_0^{\epsilon}\delta
f_\delta(\delta\mid
\delta < \epsilon)d\delta,\\
&=&\epsilon^{-({N_t}-1)}\int_0^\epsilon \delta^Ld\delta,\\
&=&\frac{{N_t}-1}{{N_t}}\epsilon. \label{Eq:EpsMean}
\end{eqnarray}
From \eqref{Eq:PwrMean} and \eqref{Eq:EpsMean},
\begin{equation}
    0\leq E[\rho\delta] < ({N_t}-1)\epsilon +
    \frac{{N_t}-1}{{N_t}}\gamma\epsilon. \label{Eq:MeanBnd}
\end{equation}
From \eqref{Eq:EpsTh} and $\phi < 1$,
\begin{equation}
\lim_{U\rightarrow\infty}\epsilon = 0. \label{Eq:Lim1}
\end{equation}
Moreover, from \eqref{Eq:EpsTh} and \eqref{Eq:GamTh},
\begin{equation}
    \gamma\epsilon =
    U^{\frac{\phi-1}{{N_t}-1}}\left(\log_2U\right)^{1-\frac{\phi\lambda}{{N_t}-1}}\left(1-\lambda\frac{\log_2\log_2U}{\log_2U}\right).
\end{equation}
If  $\lambda \geq {N_t}-1$, it follows that
\begin{equation}
\lim_{U\rightarrow\infty} E[\gamma\epsilon] = 0,\quad \text{if}\
\lambda \geq {N_t}-1.\label{Eq:Lim2}
\end{equation}
Combining \eqref{Eq:MeanBnd}, \eqref{Eq:Lim1} and \eqref{Eq:Lim2}
completes the proof.

\subsection{Proof of Theorem Sum Capacity}\label{APP:CapOrder}
The lower and upper bounds of the asymptotic sum capacity given in
\eqref{Eq:SumRate} are proved in Section~\ref{APP:CapLwBnd} and
Section~\ref{APP:CapUpBnd}, respectively.

\subsubsection{Lower Bound for Asymptotic Sum
Capacity}\label{APP:CapLwBnd}
\begin{equation}
m^\star = \arg\max_{1\leq m\leq
M}\suml{n=1}{N_t}1\{\mathcal{I}_{m,n}\neq \emptyset \}.
\end{equation}

It follows that
\begin{equation}
K^{DL}_{\max} = \suml{n=1}{N_t}1\{\mathcal{I}_{m^\star,n}\neq
\emptyset \}.
\end{equation}
From \eqref{Eq:SumRateExp}:
\begin{eqnarray}
    \mathcal{C}&\geq&E\left[\suml{n=1}{N_t}\log_2(1+
    \max_{u \in \mathcal{I}_{m^\star,n}}\SINR_u)\right],\\
    &\geq& E\left[\suml{n=1}{N_t}\log_2\left(1+
    \frac{1}{|\mathcal{I}_{m^\star,n}|}\sum_{u\in\mathcal{I}_{m^\star,n}}\SINR_u\right)\right].\label{Eq:CapOrdIneq1}
    \end{eqnarray}
By using the definition in \eqref{Eq:PrNoFb} and expanding the
expectation in \eqref{Eq:CapOrdIneq1},
\begin{eqnarray}
\mathcal{C}    &\geq& E\left[\suml{n=1}{N_t}\log_2\left(1+
    \frac{1}{|\mathcal{I}_{m^\star,n}|}\sum_{u\in\mathcal{I}_{m^\star,n}}\SINR_u\right) \mid
    K^{DL}_{\max}=N_t\right](1-P_\beta),\\
    &&+E\left[\suml{n=1}{N_t}\log_2\left(1+
    \frac{1}{|\mathcal{I}_{m^\star,n}|}\sum_{u\in\mathcal{I}_{m^\star,n}}\SINR_u\right) \mid
    K^{DL}_{\max}<N_t\right]P_\beta,\\
    &\geq& E\left[\suml{n=1}{N_t}\log_2\left(1+
    \frac{1}{|\mathcal{I}_{m^\star,n}|}\sum_{u\in\mathcal{I}_{m^\star,n}}\SINR_u\right) \mid
    K^{DL}_{\max}=N_t\right](1-P_\beta).\label{Eq:CapOrdIneq2}
\end{eqnarray}
Since the function $\log_2(\cdot)$ is convex, it follows from
\eqref{Eq:CapOrdIneq2} that
\begin{equation}
\mathcal{C}    \geq \suml{n=1}{N_t}\log_2\left(1+
    E\left[\frac{1}{|\mathcal{I}_{m^\star,n}|}\sum_{u\in\mathcal{I}_{m^\star,n}}\SINR_u\mid
    K^{DL}_{\max}=N_t\right]\right) (1-P_\beta).\label{Eq:CapOrdIneq3}
\end{equation}
By substituting \eqref{Eq:SINR} into \eqref{Eq:CapOrdIneq3},
\begin{equation}
\mathcal{C}    \geq \suml{n=1}{N_t}\log_2\left(1+
    E\left[\frac{1}{|\mathcal{I}_{m^\star,n}|}\sum_{u\in\mathcal{I}_{m^\star,n}}\frac{1+P\rho_u}{1+P\rho_u\delta_u}\mid
    K^{DL}_{\max}=N_t\right]\right) (1-P_\beta).\label{Eq:CapOrdIneq4}
\end{equation}
To simplify the above expression, we use the fact that the sequences
$\{\rho_u\}_{u=1}^U$ and $\{\delta_u\}_{u=1}^U$ are i.i.d.,
respectively. Let $\SINR_0$ denote a random variable having the same
distribution as each member of the sequence $\SINR_u$. Then
\eqref{Eq:CapOrdIneq3} can be re-written as
\begin{eqnarray}
\mathcal{C}  &=& (1-P_\beta)N_t\log_2\left(
    E\left[\frac{1+P\rho_1}{1+P\rho_1\delta_1}\mid \rho_1\geq \gamma, \delta_1\leq \epsilon\right]\right) \\
    &\geq&(1-P_\beta)N_t\log_2\left((1+P\gamma)
    E\left[\frac{1}{1+P\rho_1\delta_1 }\mid\rho_1\geq \gamma, \delta_1\leq
    \epsilon\right]\right).
    \end{eqnarray}
Since the function $\frac{1}{x}$ for $x> 0$ is convex, it follows
from the above inequality that
\begin{equation}
\mathcal{C}    \geq(1-P_\beta)N_t\left\{\log_2(\gamma) + \log_2(P) -
\log_2(1+PE[\rho_1\delta_1 \mid \rho_1\geq \gamma, \delta_1\leq
    \epsilon])\right\}.
    \end{equation}
By substituting \eqref{Eq:GamTh},
    \begin{eqnarray}
\mathcal{C}    &\geq& (1-P_\beta)N_t\left[\log_2\log_2U +
\log_2\left(1-\frac{\lambda\log_2\log_2U}{\log_2U}\right) +
    \log_2(P)-\right.\\
    &&\left. \log_2(1+PE[\rho\delta])\right].
\end{eqnarray}
Therefore,
\begin{equation}
\lim_{U\rightarrow\infty}\frac{\mathcal{C}_U}{N_t\log_2\log_2U} \geq
(1-P_\beta)[1 + \Pi_1 + \Pi_2 + \Pi_3]\label{Eq:CapOrdIneq5}
\end{equation}
where
\begin{eqnarray}
\Pi_1&=&
\lim_{U\rightarrow\infty}\frac{1}{\log_2\log_2U}\log_2\left(1-\frac{\lambda\log_2\log_2U}{\log_2U}\right),\\
\Pi_2 &=&\lim_{U\rightarrow\infty}\frac{\log_2(P)}{\log_2\log_2U },\\
\Pi_3 &=& \lim_{U\rightarrow\infty}\frac{1}{\log_2\log_2U}
\log_2(1+PE[\rho\delta]).
\end{eqnarray}
The values of $\Pi_1$ and $\Pi_2$ can be observed to be zeros. The
value of $\Pi_3$ is also equal to zero by using
Lemma~\ref{Lem:MeanProd}. Therefore, we obtain from
\eqref{Eq:CapOrdIneq5} that
\begin{equation}
\lim_{U\rightarrow\infty}\frac{\mathcal{C}_U}{N_t\log_2\log_2U} \geq
(1-P_\beta)\label{Eq:CapOrdIneq6}.
\end{equation}
By applying Lemma~\ref{Lem:PrUsrShort}, we obtain the lower bound of
the asymptotic sum capacity in \eqref{Eq:SumRate}.

\subsubsection{Upper Bound for Asymptotic Sum Capacity}\label{APP:CapUpBnd}
We can bound the sum capacity for OSDMA-TF by that for the case of
feedback from all users, denoted as $\mathcal{C}^+$. Therefore,
\begin{equation}
    \mathcal{C} \leq \mathcal{C}^+=E\left[\max_{m=1,\cdots,M}\suml{n=1}{N_t}\log_2(1+
    \max_{1\leq u \leq U}\SINR_u)\right].\label{Eq:CapUpBnd}
\end{equation}
Note the difference in the subscript for the second ``max" operator
in the above equation from that of $\mathcal{C}$ in
\eqref{Eq:SumRateExp}. The case of feedback from all users is
analyzed in \cite{Huang:JointBeamScheduleLimtFb:06}. Theorem~1 of
\cite{Huang:JointBeamScheduleLimtFb:06} shows that
\begin{equation}
N_t\log_2\log_2U \leq \mathcal{C}^+\leq N_t\log_2\log_2(UM).
\end{equation}
Therefore,
\begin{equation}
    \lim_{U\rightarrow\infty}\frac{\mathcal{C}^+ }{N_t\log_2\log_2U
    }=1.\label{Eq:CapBndLmt}
\end{equation}
From \eqref{Eq:CapUpBnd} and \eqref{Eq:CapBndLmt}, we obtain the
upper bound of the asymptotic sum capacity in \eqref{Eq:SumRate}.
Thereby, we completes the proof of Theorem~\ref{Theo:CapOrder}.

\bibliographystyle{ieeetr}


\end{document}